\documentclass[12pt]{iopart}

\usepackage{graphicx} 
\usepackage{dcolumn} 
\usepackage{bm} 

\def\sqrtsNN{\mbox{$\sqrt{s_\mathrm{NN}}$}}

\newcommand{ \be }{\begin{equation}}    
\newcommand{ \ee }{\end{equation}}    
\newcommand{ \bea }{\begin{eqnarray}}    
\newcommand{ \eea }{\end{eqnarray}}

\usepackage{color}

\begin{document}       

\begin{flushright}    
\end{flushright}

\title{Collective Dynamics at RHIC}


\author{A.H. Tang}
\address{Physics Department, P.O. Box 5000, Brookhaven National Laboratory, 
Upton, NY~11973, aihong@bnl.gov}


\date{\today}
\begin{abstract}
The property of the ``perfect liquid'' created at RHIC is probed with 
anisotropic flow measurements. Different initial conditions and their 
consequences on flow measurements are discussed. The collectivity is shown 
to be achieved fast and early. The thermalization is investigated with the 
ratio of $v_4/v_2^2$. Measurements from three sectors of soft physics 
(HBT, flow and strangeness) are shown to have a simple, linear, length 
scaling. Directed flow is found to be independent of system size. 
\end{abstract}


\section{Introduction: the perfect liquid}

\begin{figure}
\vspace{-0.2cm}
  \begin{center}
\resizebox*{8cm}{5cm}{
\includegraphics{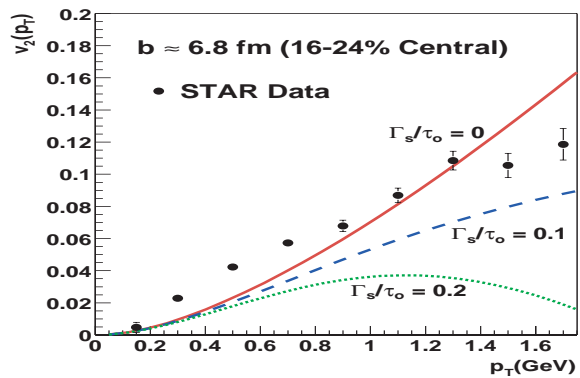}}
  \end{center}
\vspace{-0.5cm}
    \caption{Elliptic flow $v_2$ as a function of $p_T$ for different values of $\Gamma_s / \tau_o$. The data points are four-particle cumulant data from the STAR Collaboration\cite{STARCumuV2}. The difference between the ideal and viscous curves is linearly proportional to $\Gamma_s / \tau_o$. This plot is from\cite{Teaney}. \label{fig:viscosity}}
\end{figure}

As the world's first heavy ion collider, RHIC has initiated new opportunities 
for studying nuclear matter under extreme conditions. After six years of 
successful operations, the discovery of the existence of a perfect 
liquid in ultra-relativistic heavy ion collisions was 
announced\cite{PerfectLiquid}. Indications of liquid-like behavior of the matter that 
RHIC has created came in the form of large elliptic flow. Because of the 
pressure developed early in the collision, the initial spatial deformation 
due to geometry, which is quantified by eccentricity ($\epsilon$), is 
converted into the asymmetry in the momentum space, which is quantified by  
elliptic flow ($v_2$)\cite{OllitraultAndSergeiArt}. This conversion process is directly related to the 
thermalization, equation of state, etc. The wealth of data collected and 
analyzed in many aspects, including but not limited to elliptic flow, 
indicates that central Au+Au collisions can be well described by ideal 
Hydrodynamics\cite{WhitePapers}. It suggests that particles in the medium 
interact with one another rather strongly, which surprised many theoretists 
who had anticipated an almost ideal, weakly interacting gas. What is more 
interesting is that, this liquid has little viscosity and acts like a perfect 
one\cite{Teaney}. This is shown in Fig. \ref{fig:viscosity}, in which
$v_2$ from data as a function of transverse momentum ($p_T$) is compared to 
the calculation with sound attenuation length ($\Gamma_s$) scaled by the time
scale of the expansion $\tau_o$. The sound attenuation length is 
related to the shear viscosity ($\eta$) by $\Gamma_s = \frac{4}{3} \eta (e+p) $, where $e$ and $p$ are energy density and pressure, respectively.
We can see that as expected, viscosity reduces $v_2$. The calculation shows
that in order to explain the large $v_2$ observed at RHIC, one has to 
assume that the medium has an extremely small viscosity -- the 
characteristic feature of a perfect liquid.

\section{The initial condition}
\begin{figure}
  \begin{center}
\resizebox*{7.9cm}{5cm}{
\includegraphics{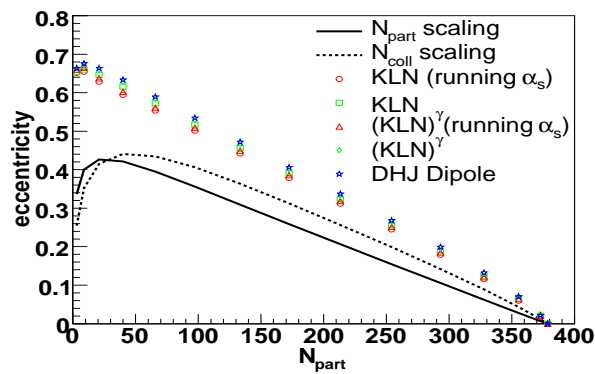}}
  \end{center}
\vspace{-0.5cm}
    \caption{Initial spatial eccentricity $\epsilon$ at midrapidity as a function of the number participants for 200 GeV Au+Au collisions from various CGC models (see\cite{CGCEcc} for the detail description of CGC models). For comparison, the initial conditions where the initial parton density at midrapidity scales with the transverse density of wounded nucleons (full line) and of binary collisions (dotted line) are also shown. This plot is from\cite{CGCEcc}  \label{fig:CGC_ecc}}
\end{figure}

\begin{figure}
\vspace{-0.2cm}
  \begin{center}
    \begin{minipage}[t]{0.48\linewidth}
\resizebox*{9cm}{5cm}{
\includegraphics{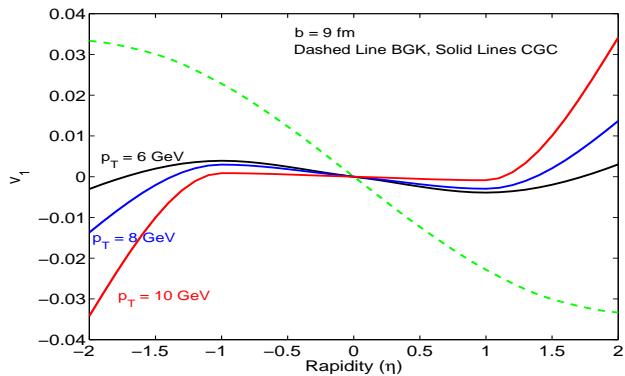}}
    \end{minipage}\hfill
    \begin{minipage}[t]{0.48\linewidth}
    \vspace{-5.2cm}  \caption{The directed flow $v_1$ as a function of $\eta$ for different $p_T$ at $b=9$ fm. Both the CGC model and BGK model are given for comparison. This plot is from[]. \label{fig:CGC_v1}}
    \end{minipage}
  \end{center}
\vspace{-1.cm}
\end{figure}

\begin{figure}[ht]
\vspace{-0.5cm}
\begin{center}
\resizebox{
\textwidth}{!}{
\resizebox*{10cm}{7cm}{
\includegraphics{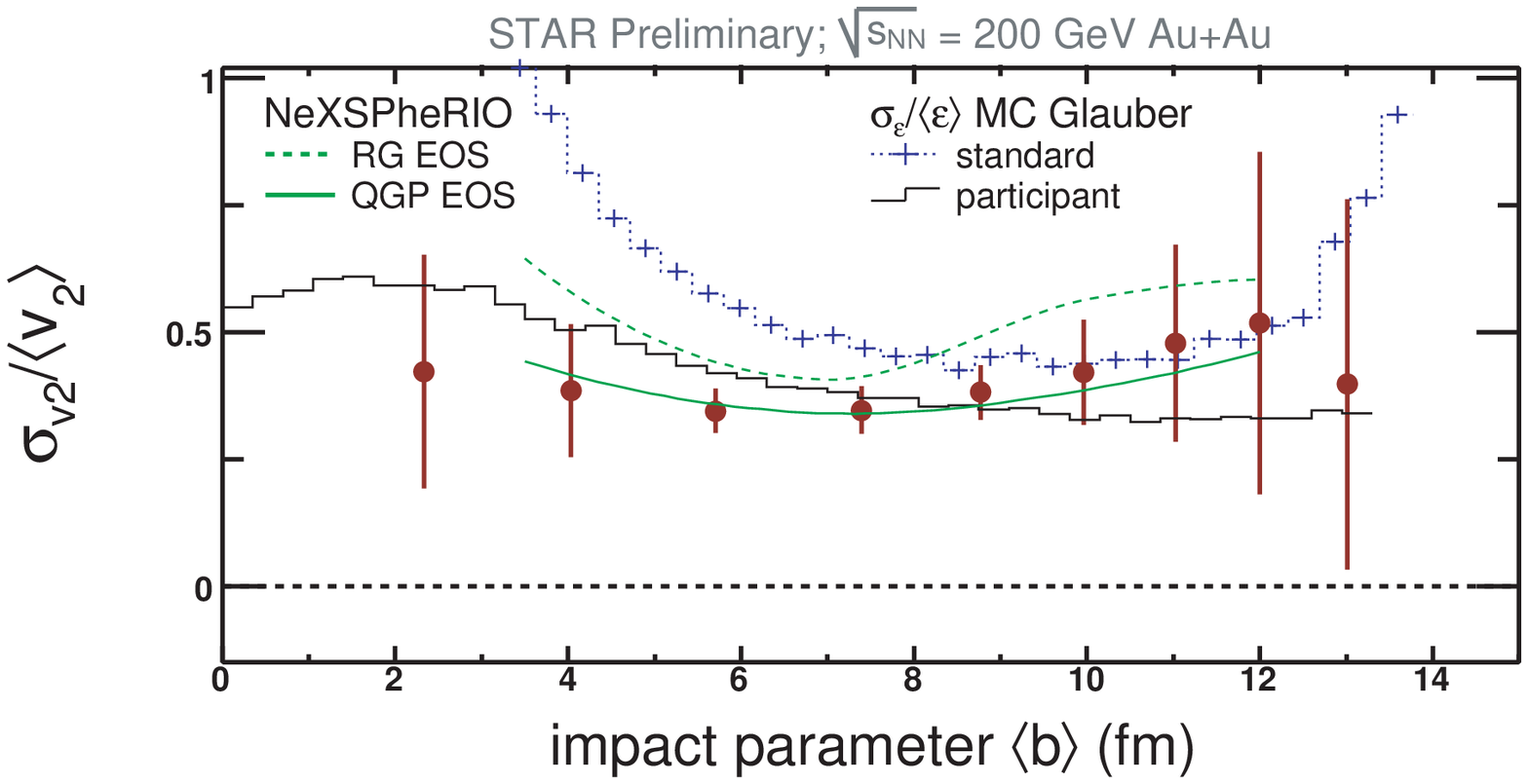}}
\resizebox*{10cm}{6.6cm}{
\includegraphics{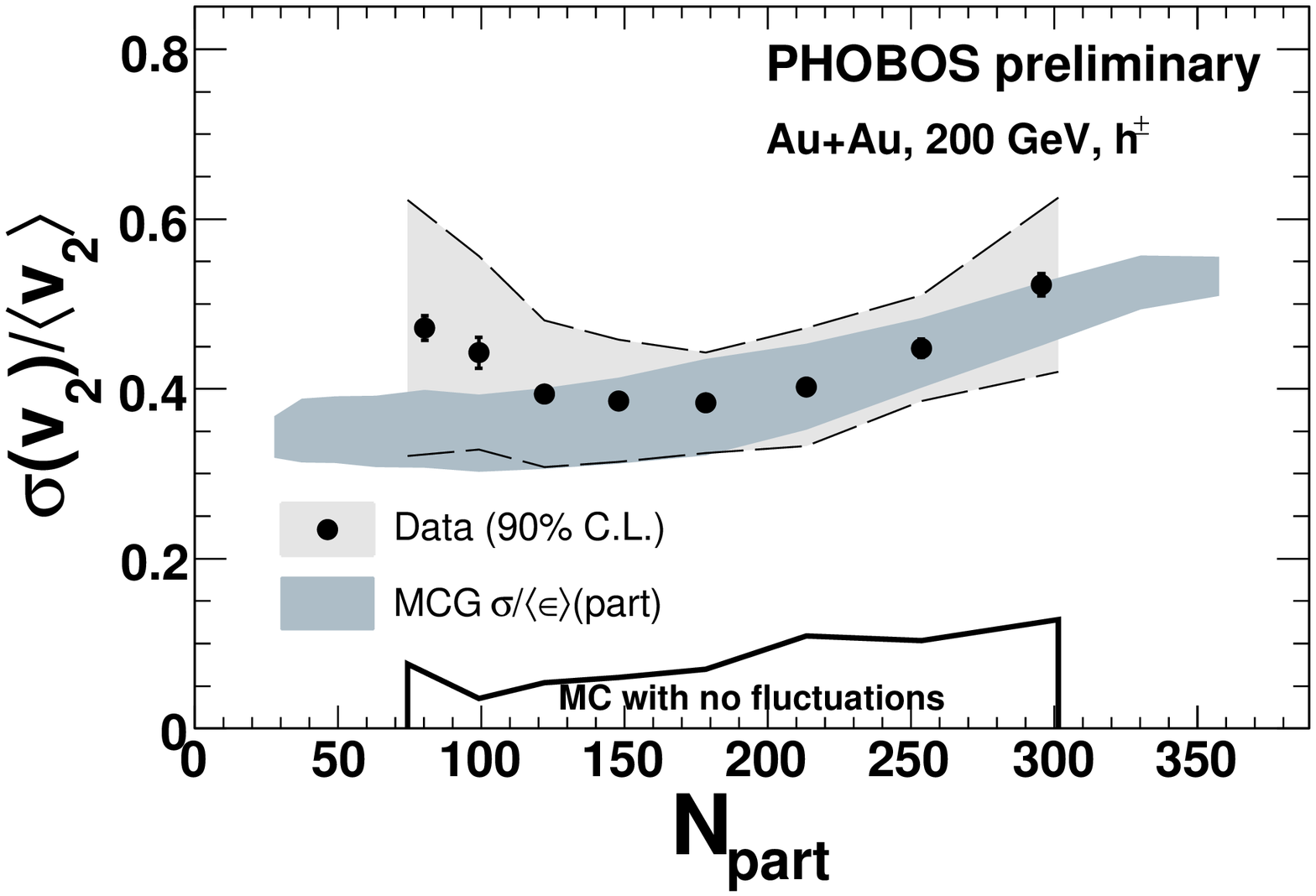}}}
\caption{The r.m.s. width of the $v_2$ distribution ($\sigma_{v2}$) scaled by the mean $v_2$. Data are presented versus impact parameter (left by STAR) and number of participants (right by PHOBOS). In the left plot, together shown are eccentricity fluctuations $\sigma/\langle\epsilon\rangle$ calculated from the Monte-Carlo Glauber model with standard eccentricity (crosses) and participant eccentricity (step-line), the latter calculation is also done by PHENEX (dark contour in the right plot) \label{fig:v2Fluct}}
\end{center}
\vspace{-0.5cm}
\end{figure}

The viscosity is so small that the initial spatial eccentricity 
is converted to momentum anisotropy with a high efficiency, and this
process results in large amount of $v_2$ as reported by RHIC experiments. 
In this explanation one assumes that the initial 
spatial eccentricity is from Glauber source\cite{Glauber}. Recent theoretical 
work (Fig. \ref{fig:CGC_ecc}) shows that a different initial condition like 
Color Glass Condensate (CGC) will give a much larger initial spatial 
eccentricity than that is from Glauber source. As a consequence of that, 
the viscosity has to be finite, as opposed to the close-to-zero viscosity in a 
perfect liquid, in order to reduce the $v_2$ to the level that matches 
the data. 
Thus the matter that RHIC has created can be explained either by a perfect 
liquid with a Glauber source or, a viscous matter with a CGC source. To 
distinguish between the two, one has to understand the initial condition. 
However it is not easy to trace the initial condition, because with it the 
system starts, and after that the system has gone through thermalization, 
a possible QGP phase, hadronic interactions and freeze out. A lot of early 
information can be easily washed out or completely lost due to various 
effects at later stages. Nevertheless, both theoretists and experimentalists 
begin to realize the importance of the initial condition, and 
starts to trace its footprints. Fig. \ref{fig:CGC_v1} shows that
for high $p_T$ particles the $v_1$ (solid lines) from CGC flips sign 
at $\eta \simeq 1.2$, and becomes positive for higher values of rapidity. 
That means particles are flowing in the same direction as the projective 
spectator. In the conventional factorized jet production(dashed line),
the high $p_T$ $v_1$ is negative and in the same direction as the low $p_T$ 
bulk directed flow. It would be interesting for experimentalists to test
this novel prediction from CGC in the future. One can also exam 
the initial condition by studying the fluctuation of elliptic flow. Both 
STAR\cite{SorensenQM06} and PHOBOS\cite{LoizidesQM06} collaboration has 
measured (see Fig.\ref{fig:v2Fluct}) the $v_2$ fluctuation and compared it 
to the fluctuation from initial conditions assuming Glauber sources.
The $v_2$ fluctuation is found to be significant ($\sim 40\%$ relatively), 
and most of it can be explained by the fluctuation from the Glauber model
as the initial condition. It means that, again, the conversion process from 
the initial spatial eccentricity to momentum anisotropy is so complete
that little room is left for fluctuations of other dynamic processes.

\section{Collectivity and thermalization}
\begin{figure}
\vspace{-0.2cm}
  \begin{center}
    \begin{minipage}[t]{0.48\linewidth}
\resizebox*{9cm}{5cm}{
\includegraphics{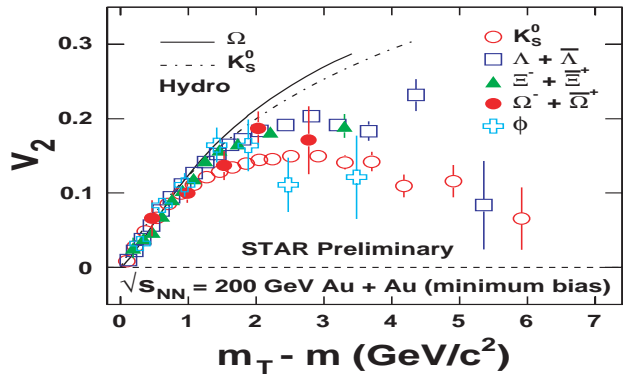}}
    \end{minipage}\hfill
    \begin{minipage}[t]{0.48\linewidth}
    \vspace{-5.3cm}  \caption{$v_2$ as a function of transverse kinetic energy for Multi-Strange baryons. Also shown is the Hydro calculation for $\Omega$ and $k_s^o$. The data points are from \cite{Lu}. \label{fig:MSHv2}}
    \end{minipage}
  \end{center}
\vspace{-1.cm}
\end{figure}

After the initial collision, particles begin to exchange momentum and the 
system begins to build up collectivity. Knowing when and how the 
collectivity is achieved is the first step towards 
understanding the dynamics in a hot and dense environment.
This can be addressed by studying the $v_2$ of $\phi$ and $\Omega$. Both of 
them are expected to have small hadronic cross section\cite{TeaneyHydro} 
thus are less affected by hadronic interactions. The other reason to choose 
$\phi$ for this purpose is because of its long lifetime -- it decays outside 
of the fireball and is not formed by $k^+ k^-$ coalescence, thus it picks up 
little information from a later stage. Fig. \ref{fig:MSHv2} shows that, 
although $\phi$ and $\Omega$ tends to suffer much less rescatterings in the 
hadronic stage of the collision, their $v_2$ are found to be as high as other 
hadrons at a given $p_T$. Hence the collectivity must be developed fast 
at a pre-hardonic stage. 

\begin{figure}
\vspace{-0.2cm}
  \begin{center}
    \begin{minipage}[t]{0.48\linewidth}
\resizebox*{9cm}{5cm}{
\includegraphics{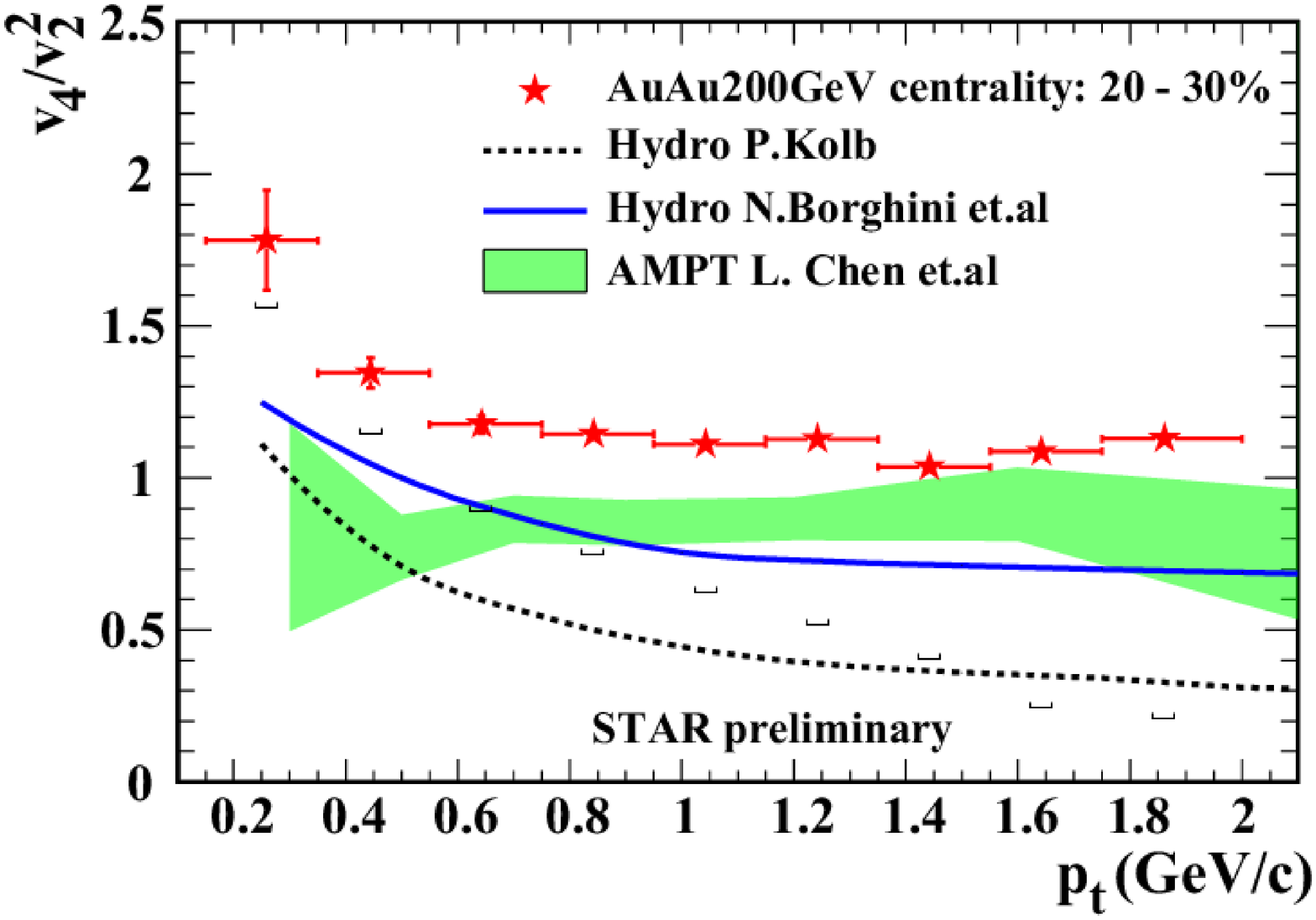}}
    \end{minipage}\hfill
    \begin{minipage}[t]{0.48\linewidth}
    \vspace{-5.3cm}  \caption{$v_4/v_2^2$ as a function of $p_T$. $v_4$ is measured by the three-particle cumulant method, and $v_2$ is measured by the four-particle cumulant method. Also shown are model calculations. This figure is from \cite{Bai}\label{fig:v4v2SqrRatio}}
    \end{minipage}
  \end{center}
\vspace{-0.3cm}
\end{figure}

Building up collectivity does not necessarily mean that the system is 
thermalized. In order for the system to be thermalized, particles in the 
system have to ``talk'' to each other intensively so that the information 
like the initial spatial anisotropy can be passed on to all particles. This 
process depends on number of collisions encountered by each
particle. It is expected that both $v_2$ and $v_4$ are proportional
to the number of collisions per particle, and thus the ratio of $v_4/v_2^2$
decreases with it\cite{Borghini}. In Fig.\ref{fig:v4v2SqrRatio}, this ratio is 
plotted against $p_T$ and compared to theoretical calculations. The Hydro 
calculation done by Borghini and Jean-Yves\cite{Borghini} suggests that in 
ideal hydrodynamics, this ratio decreases as a function of $p_T$. Another 
version of Hydrodynamic calculation\cite{Peter} shows a similar trend 
with smaller magnitude. The calculation from the AMPT\cite{Chen} model shows 
a more or less flat shape. The data points are higher than theoretical 
calculations but the systematical errors are also large. It is desirable 
that in the future the uncertainty from both experiment measurement and 
theoretical calculation can be reduced, so that the degree of thermalization 
can be tested.

\section{Scaling of soft physics}
\begin{figure}
\vspace{-0.5cm}
  \begin{center}
    \begin{minipage}[t]{0.48\linewidth}
\resizebox*{9.5cm}{9cm}{
\includegraphics{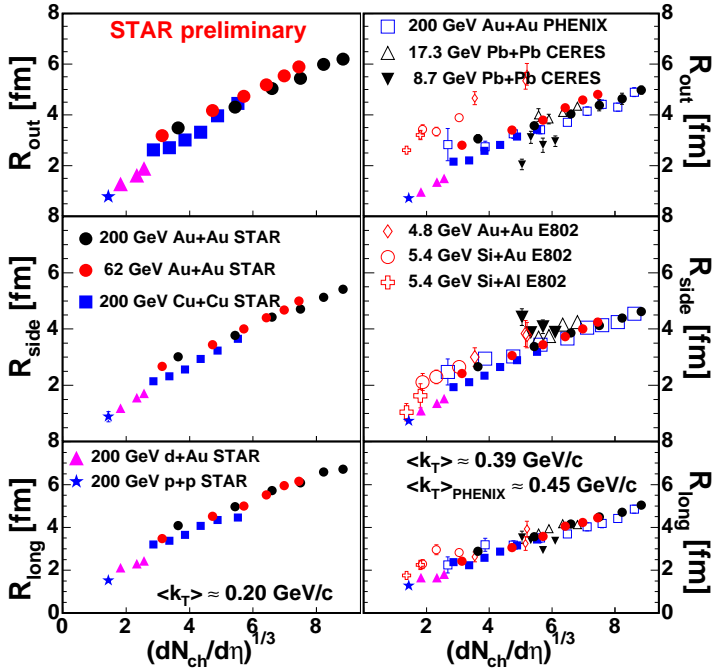}}
    \end{minipage}\hfill
    \begin{minipage}[t]{0.48\linewidth}
    \vspace{-9cm}  \caption{Femtoscopic radii dependence on the number of charged particles. Left panel: results from STAR experiment only for $<k_T> \simeq$0.20 GeV/c; right panel: STAR results combined with data from PHENIX,CERES and E802 experiments, mean value of $k_T$ is given on the plot. This plot is from \cite{Chajecki}.\label{fig:hbt}}
    \end{minipage}
  \end{center}
\vspace{-1.cm}
\end{figure}

\begin{figure}
  \begin{center}
    \begin{minipage}[t]{0.48\linewidth}
\resizebox*{9.5cm}{7.5cm}{
\includegraphics{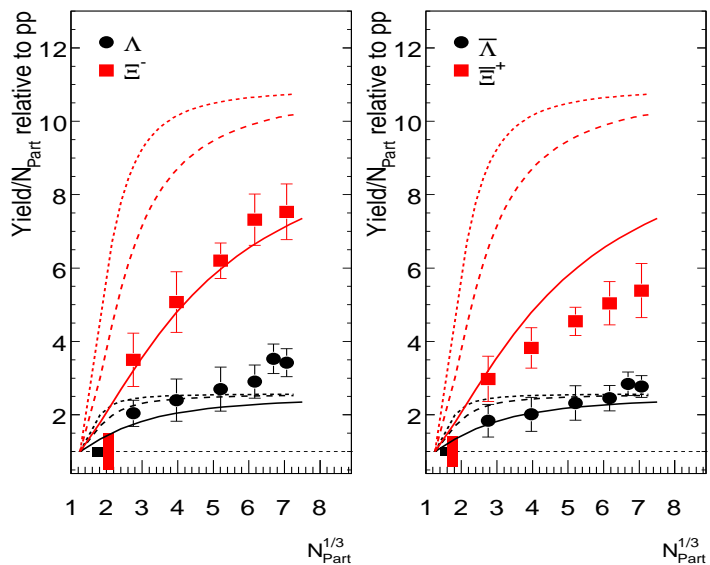}}
    \end{minipage}\hfill
    \begin{minipage}[t]{0.48\linewidth}
    \vspace{-8cm}  \caption{Strangeness enhancement as a function of $p_T$. Also shown are three theory curves which represent the evolution with collision participants ($N_{part}$) of the expected enhancement factors. The correlation volume for strangeness enhancement is calculated as $V = (N_{part}/2)^\alpha V_o$, where $V_o = 4/3.\pi R^3$ and $R$ is the radius of the proton. The three curves correspond to values of $\alpha$ of 1(short dashed line), 2/3 (long dashed line) and 1/2(solid line) respectively. The figure is re-plotted based on Fig.2 in \cite{Lamont}.\label{fig:strangeness}}
    \end{minipage}
  \end{center}
\vspace{-0.3cm}
\end{figure}

\begin{figure}
\vspace{0.5cm}
  \begin{center}
    \begin{minipage}[t]{0.48\linewidth}
\resizebox*{8cm}{5.5cm}{
\includegraphics{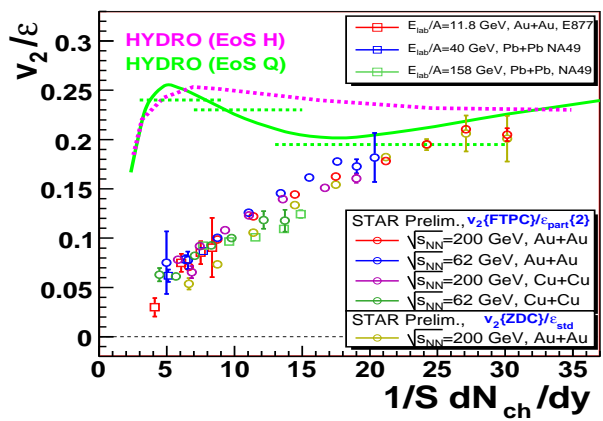}}
    \end{minipage}\hfill
    \begin{minipage}[t]{0.48\linewidth}
    \vspace{-5.85cm}  \caption{$v_2/\epsilon$ vs. $1/S dN/dy$ measured by the STAR Collaboration. Where $S$ is the overlap area and $dN/dy$ is the yield for charged particles at midrapidity. This plot is from\cite{Sergei}. \label{fig:STARv2ecc}}
    \end{minipage}
  \end{center}
\vspace{-0.5cm}
\end{figure}

\begin{figure}
  \begin{center}
    \begin{minipage}[t]{0.48\linewidth}
\resizebox*{8.2cm}{5cm}{
\includegraphics{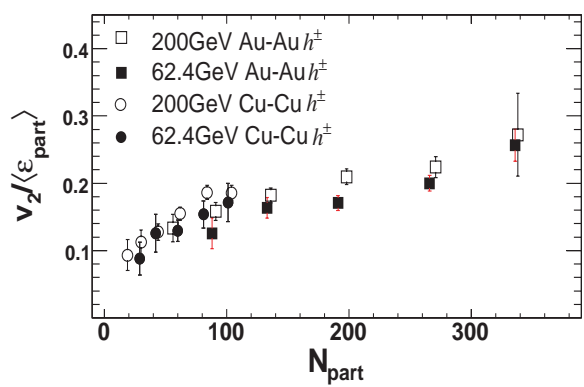}}
    \end{minipage}\hfill
    \begin{minipage}[t]{0.48\linewidth}
    \vspace{-5.3cm}  \caption{$v_2$ vs. $N_{part}$, for Cu+Cu and Au+Au collisions at $\sqrtsNN = 62.4$ and $200$ GeV. 1-$\sigma$ statistical error bars are shown.This plot is from\cite{PHOBOS}\label{fig:PHOBOSv2ecc}}
    \end{minipage}
  \end{center}
\vspace{-1.cm}
\end{figure}

The number of collisions encountered by each particle on its way out
not only plays an important role in thermalization, but also
leads to a simple, but interesting scaling of soft physics. Fig. \ref{fig:hbt}
shows that for different collision energies and over a wide range of collision
systems, the HBT radii show a nice linearity if plotted against $dN/dy^{1/3}$,
 which is proportional to the source's length, and in turn 
relates to the number of interactions for a particle on its way out. 
$R_{out}$ is an exception because it includes both space and time information 
thus the simple scaling with length is not expected. A similar $dN/dy^{1/3}$ 
scaling is also observed\cite{Lamont} in the strangeness yield relative to pp. 
Fig. \ref{fig:strangeness} shows a good linearity if the relative yield 
of $\Omega$ and $\Xi$ are plotted as a function of $dN/dy^{1/3}$. Also shown 
in the figure is the theoretical calculation of the enhancement with the 
correlation volume $V = (N_{part}/2)^{\alpha}V_o$, where $V_o=4/3.\pi R^3$ 
and $R$ is the radius of the proton. The curve which fits the shape of the 
data the best is for the case of $\alpha=1/3$, which indicates that  
length plays an important role in strangeness production.
Such linearity 
can be seen in flow measurements as well. In Fig. \ref{fig:STARv2ecc}, the
$v_2$ is scaled by the initial eccentricity and plotted as a function of 
paticle's density $1/S dN/dY$, which is also
proportional to the length of the system because 
$dN/dY$ is proportional to the volume and $S$ is the overlap area. 
Over a broad range of collision 
energies and system sizes, we observe a good linear relationship between 
$v_2/\epsilon$ and $1/S dN/dY$. This linear relation disapears if the same 
quantity plotted against $N_{part}$ (Fig. \ref{fig:PHOBOSv2ecc}), which is 
directly related to the volume.

The simple linear scaling from three important sectors of soft physics 
(HBT, strangeness, flow) suggests that the number of collisions encountered by 
each particle plays an important role in soft physics. One may 
venture\cite{Helen} to predict $v_2$, HBT radii and the relative strangeness 
yield based on this simple scaling, without knowing anything about the 
collision(energy, system size etc.).

\section{Directed flow}
\begin{figure}[ht]
\begin{center}
\resizebox{
\textwidth}{!}{
\resizebox*{10cm}{6.6cm}{
\includegraphics{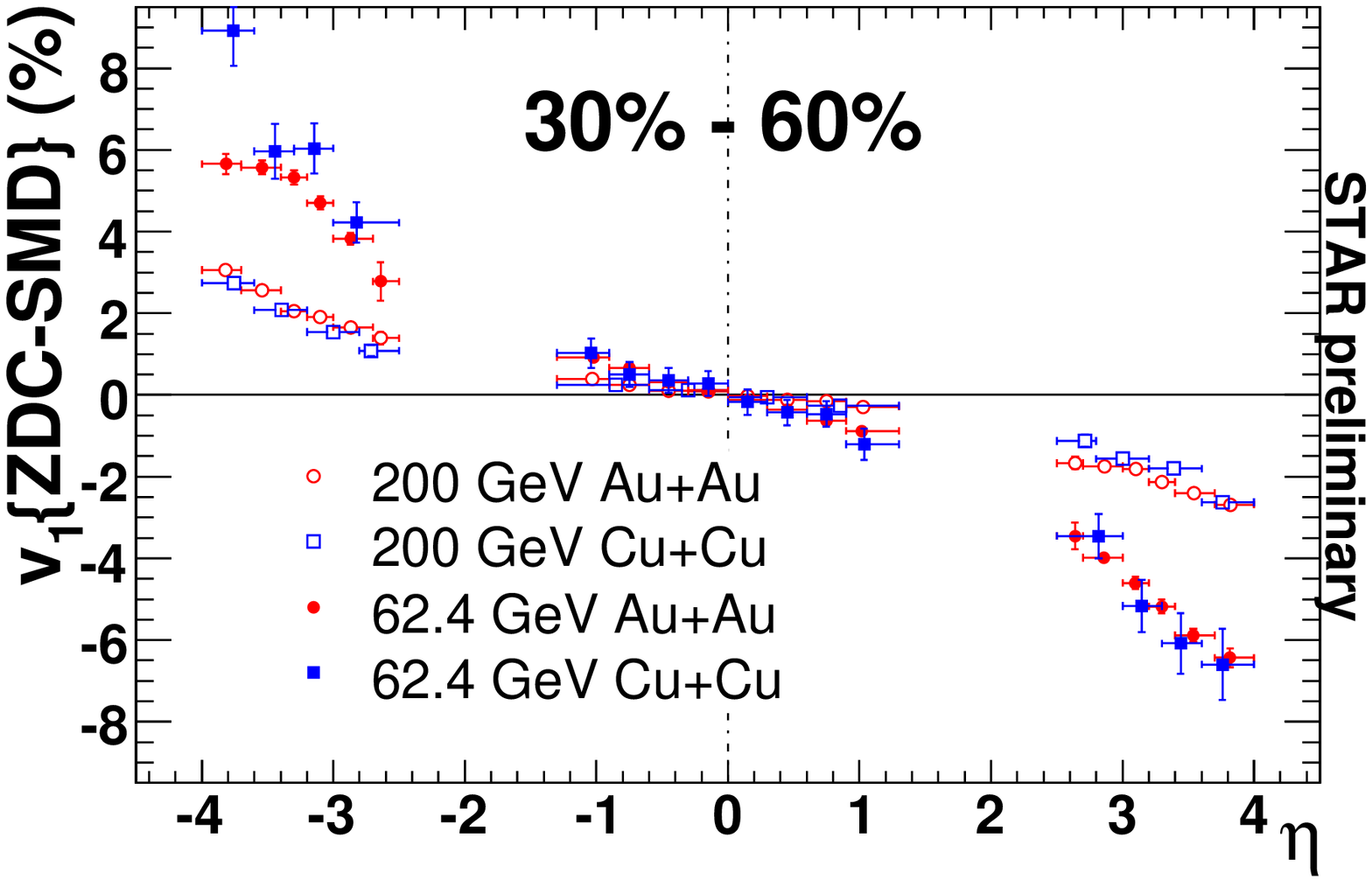}}
\resizebox*{10cm}{6.8cm}{
\includegraphics{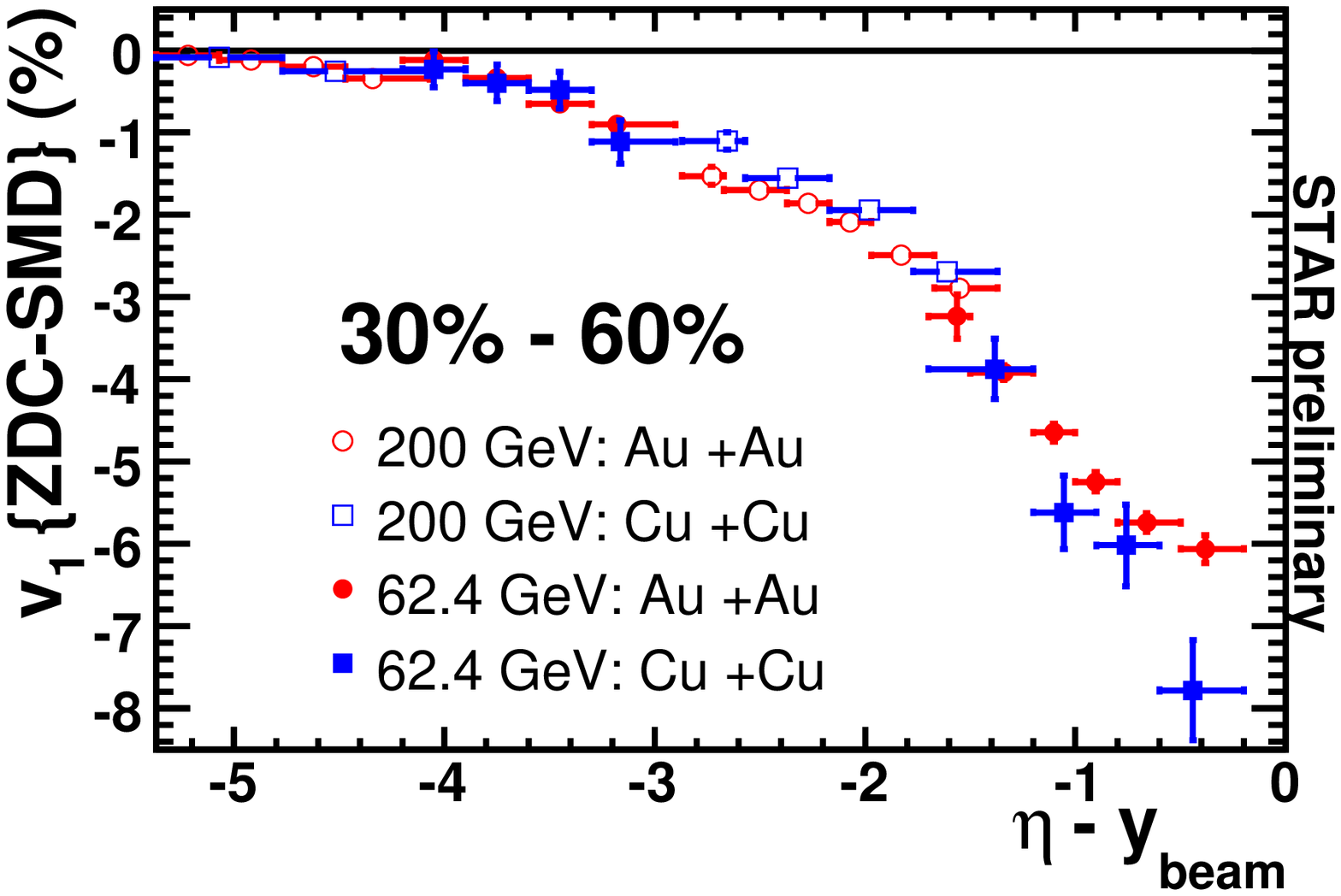}}}
\caption{Left: Charged-hardon $v_1$ vs. $\eta$, for Au+Au and Cu+Cu collisions at  $\sqrtsNN = 62.4$ and $200$ GeV. Right: The same data but plotted as a function of $v_1$ vs. $\eta - y_{beam}$. Both plots are from \cite{Gang}. \label{fig:v1}}
\end{center}
\vspace{-0.5cm}
\end{figure}

Directed flow ($v_1$) describes the ``bounce-off'' motion of particles 
away from midrapidity. As an important tool to probe the system at forward
rapidity, it complements our understanding of the dynamics at midrapidity.
Directed flow from different energies at SPS has been studied in \cite{NA49}, 
however its system size dependence has not been well explored. $v_1$ for 
Au+Au collisions at both $\sqrtsNN = 62.4$ and $200$ GeV have been 
measured\cite{STARv1}, the Cu+Cu data that RHIC experiments collected in 
year 2005 at the same two energies gives us a good opportunity to study the 
system size dependence. The left plot in Fig. \ref{fig:v1} presents $v_1$ 
as a function of pseudorapidity measured by the STAR Collaboration. Data from 
Cu+Cu collisions and Au+Au collisions
at both energies ($\sqrtsNN = 62.4$ and $200$ GeV) are shown. The data points
fall into two bands, one is for $\sqrtsNN = 62.4$ GeV and the other one is 
for $\sqrtsNN = 200$ GeV. From Au+Au collision to Cu+Cu collision the 
system size is reduced by $1/3$, however the $v_1$ does not change. This
is true even for the region near midrapidity, where $v_2$ for Cu+Cu 
collisions is considerably lower that that for Au+Au collisions~\cite{Sergei}. 
Unlike $v_2/\epsilon$ which scales with system length, $v_1$
is found to be independent of system size. Instead, it scales with the 
incident energy. A possible explanation to the different scalings for 
$v_2/\epsilon$ and $v_1$  might comes from the way in which they are 
developed : To produce $v_2$, intensive momentum exchanges among particles are 
needed (and remember number of momentum exchanges is related to the length), 
while to produce $v_1$, one in principle needs only different rapidity losses, 
which has a connection to the incident energy, for particles having different 
distances away from the central point of the collision.

One may also test the limiting fragmentation hypothesis\cite{LimFrac}, 
which has successfully described particle's yield and flow at forward rapidity, 
with different system sizes. The right plot in Fig.\ref{fig:v1} re-plotted 
the same $v_1$ results as a function of $\eta - y_{beam}$. We can see that 
within three units from beam rapidity, most data points fall into a universal 
curve. This extends the validity of limiting fragmentation to different 
collision system sizes. There are evidences\cite{Bai} show that limiting fragmentation
also works for higher harmonics like $v_4$.

\vspace{-0.2cm}

\section{Summary}
\vspace{-0.1cm}
In summary, rich results from RHIC support a Hydrodynamic expansion of 
a thermalized fluid, in which the collectivity is achieved fast and at the 
very early time. Understanding the initial condition plays a key role in 
understanding what happens thereafter. Studying elliptic flow fluctuation,
as well as directed flow for high $p_T$ particles, may help us constraint the 
initial condition. A few key observables from soft physics are found scaling 
with system length, which is directly related to the average number of 
interactions for a particle on its way out. Directed flow is found to 
depend on the incident energy but not on the system size. Limiting 
fragmentation holds for different collision energies, systems and flow 
harmonics.

\end{document}